# Very Deep Super-Resolution of Remotely Sensed Images with Mean Square Error and Var-norm Estimators as Loss Functions

Antigoni Panagiotopoulou[1], Lazaros Grammatikopoulos[2], Eleni Charou[3], Emmanuel Bratsolis[4], Nicholas Madamopoulos[5] and John Petrogonas[6]

[1]Institute of Informatics and Telecommunications, NCSR Demokritos, Neapoleos, Agia Paraskevi, 15310, Athens, Greece

[2]Department of Surveying and Geoinformatics Engineering, University of West Attica, 28 Agiou Spyridonos, Aigaleo 12243, Athens, Greece

[3]Institute of Informatics and Telecommunications, NCSR Demokritos, Neapoleos, Agia Paraskevi, 15310, Athens, Greece

[4]Department of Surveying and Geoinformatics Engineering, University of West Attica, 28 Agiou Spyridonos, Aigaleo 12243, Athens, Greece

[5]Department of Aeronautical Sciences, Helenic Air Force Academy, Acharnes 13672, Dekeleia, Greece and Department of Electrical Engineering, The City College of New York, 160 Convent Avenue, New York, NY 10031, USA

[6]Map Ltd, 6 Aristotelous, Cholargos 15562, Athens, Greece

**Abstract:** In this work, very deep super-resolution (VDSR) method is presented for improving the spatial resolution of remotely sensed (RS) images for scale factor 4. The VDSR net is re-trained with Sentinel-2 images and with drone aero orthophoto images, thus becomes RS-VDSR and Aero-VDSR, respectively. A novel loss function, the Var-norm estimator, is proposed in the regression layer of the convolutional neural network during re-training and prediction. According to numerical and optical comparisons, the proposed nets RS-VDSR and Aero-VDSR can outperform VDSR during prediction with RS images. RS-VDSR outperforms VDSR up to 3.16 dB in terms of PSNR in Sentinel-2 images.

## 1. Introduction

Remotely sensed (RS) images such as satellite and aerial images constitute a great source of information about ground and above ground material. The spatial resolution of RS images frequently needs to be increased or enhanced [1-2]. The problem of generating a high-resolution (HR) image given a low-resolution (LR) image is commonly referred to as Single Image Super-Resolution (SISR). SISR is broadly used in various computer vision applications, where increased image details are a prerequisite. Lately, learning-based techniques for image SISR are widely employed. Convolutional neural networks demonstrate improved performance and accuracy over other previous techniques [1, 3].

Very Deep Super-Resolution (VDSR) is a SISR technique that was presented in [3-4]. It is based on very deep convolutional network and presents high accuracy. Residual learning and gradient clipping are utilized to fasten convergence without the presence of exploding gradients, whereas at the same time assuring training stability. Furthermore, the VDSR net model can perform resolution enhancement by factors 2, 3, and 4, thus a multiscale super-resolution (SR) problem is addressed by a single network.

VDSR technique has been utilized as a tool for super-resolving RS images in [5-6]. In [5], a novel network called remote sensing deep residual-learning (RS-DRL) is presented for spatial resolution improvement of Sentinel-2A images by a factor of 2. The test dataset consists of B02, B03 and B04 bands. RS-DRL outperforms both bicubic interpolation and VDSR on peak signal to noise ratio (PNSR) metric. In [6], real and synthetic images are utilized to evaluate the performance of VDSR. Firstly, it is proved that VDSR can super-resolve images that have been distorted by effects like diffraction. The testing images were drawn from the dataset used in VDSR re-training, as well as from an aerial dataset. Secondly, testing images of a resolution chart lead to the conclusion that VDSR has the capability of increasing the image visual contrast, but it produces ringing effects on the SR image. However, no novel loss functions have been yet presented in the literature when VDSR being applied on RS images.

The present work newly modifies the already presented VDSR network [3] for coping with RS images. Specifically, VDSR is re-trained with RS images, Sentinel-2 and aero orthophotos, while the Var-norm estimator [7] serves as a loss function. The novelty of this work lies on the utilization of the Var-norm estimator in the regression layer during re-training, as well as during prediction. Prediction tests are performed with RS images which are partially known to the training dataset. Results prove the predominance of the proposed nets over VDSR.

This paper is organized into 5 sections. The Var-norm estimator as loss function is presented in Section 2. VDSR, RS-VDSR and Aero-VDSR nets, along with the utilized RS data, are described in Section 3. Section 4 presents the experiments which were carried out. Conclusions are drawn in Section 5.

## 2. The Var-norm Estimator as Loss Function

The mean square error (MSE) estimator was used in the regression layer of VDSR network [3] during its training with natural images. Therefore, MSE served as the loss function. In deep learning, the loss function represents the cost of inaccuracy of the network and is the distance between training images and the associated predictions. The loss function is the function that the entire training progress tries to minimize.

The loss function which is utilized in deep learning training should be a computationally feasible function. Ideally, the choice and design of the loss function will be dictated by a combination of factors, such that it satisfies the requirements of the use case or specific application. An adaptive loss function could prove very useful. In the present work, the Var-norm estimator [7] is selected to be employed in the regression layer of the network during and prediction. Comparisons are carried out with the MSE estimator as loss function.

The estimator which is utilized in the regression layer of the deep network measures the difference of the estimated HR image from the corresponding target HR image in each iteration of the training procedure. In particular, the influence function or $\psi$ −function [7-9] of the employed error norm takes as arguments the elements of the matrix resulting from the subtraction of the estimated image from the target image. Small difference of the HR images under comparison means successful HR estimation. Therefore, the employed influence function should reward the small differences between the compared images and penalize the large differences. Specifically, the influence function arguments should be given large or small weights in case of correct or incorrect HR estimate, respectively. In this way, only the correct measurements strongly affect the solution and the SR reconstruction procedure is not dominated by erroneous HR estimates. Then the utilized estimator provides for preserving fidelity to the target training images. The Var-norm estimator presents the following $\psi$ −function:

$$\psi_{var}(x) = \frac{x}{R + var(abs(F))}, \tag{1}$$

where $F$ denotes a matrix that is the difference between the estimated HR image and the corresponding target HR image. $x$ is an element of the matrix $F$ and represents estimations errors. The symbol $abs$ represents absolute value and $var$ stands for variance. $R$ is a parameter which accounts for stability of the estimator.

In case the HR image has not been correctly estimated, the elements of the matrix $F$ have large absolute values. Therefore, the variance of the estimation error, $var(abs(F))$, also presents large value. In the specific case, due to the proposed influence function (1), the values of the elements of $F$ will be decreased. Thus, the particular values denote incorrect estimation, so they will be penalized. On the contrary, in case the HR image has been correctly estimated, the introduced influence function increases the values of the elements $F$. Consequently, in each iteration of the training procedure for each estimated HR image the weight, which is given to the argument of the influence function, is automatically tuned. Fig. 1 demonstrates graphical representation of the Var-norm influence function. The particular function is a straight line. The point of intersection between the graph of the function and the y-axis is constant and equals zero. However, the line slope that equals $(R + var(abs(F)))^{-1}$ varies according to the estimation error which is represented by $F$. Hence, estimation errors ranging from small to large values make the line of the influence function change from high slope to low slope, respectively. Consequently, the influence function of the Var-norm estimator presents a variant slope which allows each measurement to characterize the bias introduced in the solution. Here, measurement stands for the absolute difference of the estimated HR image from the target HR image.

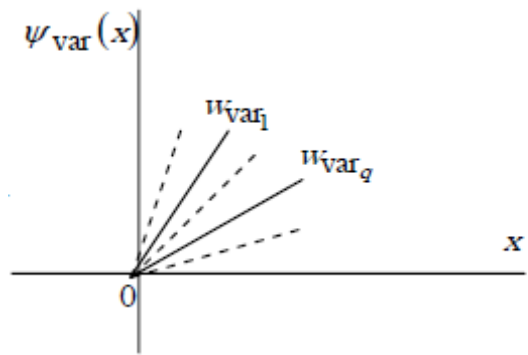

**Fig. 1.** The $\psi-$ function of the Var-norm estimator. The symbol $w_{var}$ stands for slope. The proposed Var-norm $\psi-$ function presents variant slope which allows each measurement to characterize the bias introduced in the solution. Measurement is the difference of the estimated HR image from the target HR image [7].

## 3. The Network Models VDSR, RS-VDSR and Aero-VDSR

The VDSR network has been already trained with natural images [3] with the MSE estimator in the regression layer during training. In this work, the MSE estimator is also used in the regression layer during VDSR prediction. Concerning RS-VDSR and Aero-VDSR networks, two different training/prediction methodologies are followed, namely -VDSR(MSE) and -VDSR(Var-norm). When following the -VDSR(MSE) methodology, the MSE estimator is used in the regression layer during VDSR re-training and for prediction. In the case of -VDSR(Var-norm) methodology, the Var-norm estimator is utilized in the regression layer during VDSR re-training as well as for prediction.

### 3.1. VDSR

VDSR is a SISR technique [3-4] that is based on very deep convolutional network and adopts residual learning strategy to achieve fast convergence. The VDSR network structure is depicted in Fig. 2, where pairs of convolutional and nonlinear layers are cascaded repeatedly. An interpolated low-resolution image (ILR) enters the network and after going through the layers gets transformed into a HR image. A residual image is predicted by the network and its addition to the input ILR gives the SR output or desired HR image. Each convolutional layer employs 64 filters. Gradient clipping is also utilized so that to avoid exploding gradients and to assure training stability. Furthermore, the single network VDSR can handle a multiscale SR problem, since resolution enhancement per factors 2, 3, and 4 can be performed.

 In this work, the VDSR net with 20 convolutional layers is utilized. The receptive field equals 41×41 and represents the pixels size of the input image patches that enter the network. VDSR has already been trained on natural images [3] with MSE as training loss function. Note that the RS images are completely unknown to the VDSR network.

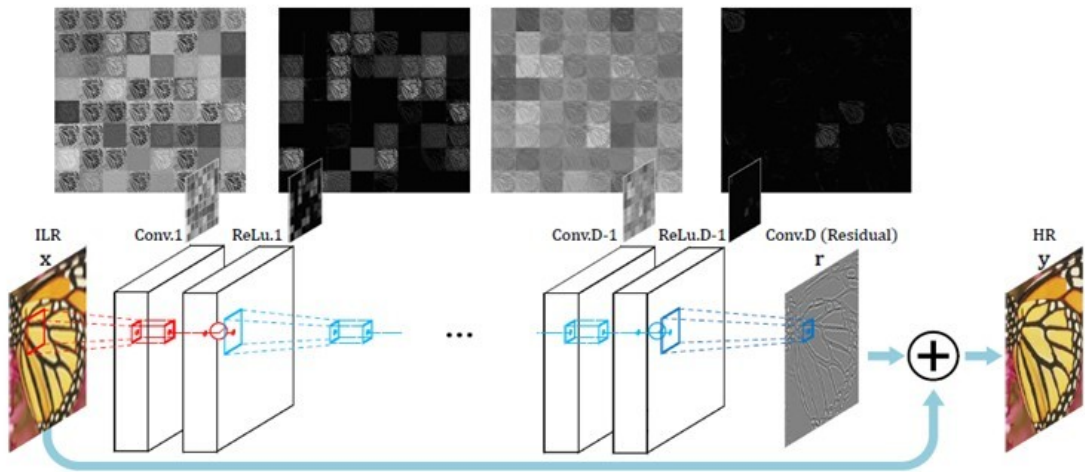

**Fig. 2.** The VDSR network structure [3].

**3.2. RS-VDSR**

The VDSR network is re-trained with Sentinel-2 images and is renamed to RS-VDSR. Multi-scale re-training per factors 2, 3, and 4 has been performed. Particularly, the dataset Demokritos [10] is utilized for the training. This is a publicly available dataset and the area of study consists of the Ionian islands in Greece, namely the islands of Corfu, Cephalonia, Zakynthos, Kithira, Meganisi, Paxoi and Lefkada as depicted in Fig. 3.

Demokritos dataset was constructed in a two steps procedure:

(a) Satellite Image Acquisition step followed by

(b) a Dataset creation step

In the Satellite Image Acquisition step, 125 Sentinel-2 tiles of the Study Area, between January 2017 and April 2019, with low cloud coverage were acquired. Since optical satellite imagery may be still contaminated with clouds and shadows, preliminary processing steps were taken to clean the data. The initial tiles were exported in a map scale of 1/30,000 through Qgis using Sentinel's Hub plugin in the following forms (a) Georeferenced TIFF images (Reference System WGS 84 Pseudo Merctor) and (b) JPG images. In the Demokritos Dataset Creation step, the obtained satellite tiles were divided into 2,925 non-overlapping image patches. Each patch is a natural color RGB image of size 227×227 pixels with spatial resolution of 10m per pixel. Agricultural and non-agricultural scenes are depicted in Demokritos dataset. The agricultural scenes include the non-irrigated arable land, permanently irrigated land, rice fields, vineyards, olive groves, pastures, annual crops associated with permanent crops, complex cultivation patterns and land principally occupied by agriculture with significant areas of natural vegetation. The non-agricultural scene class includes all the remaining CORINE Land Cover 2018 [11] classes.

The Demokritos dataset contains 2,925 images of pixels size 227×227. Each image was cut into 6 patches of 41×41 pixels, thus totally 17,550 observation patches were created and used for VDSR re-training.

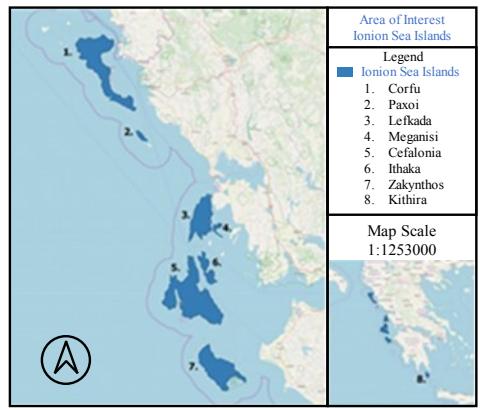

**Fig. 3.** Area which is covered by the Demokritos dataset [10].

### 3.3. Aero-VDSR

The VDSR network is multi-scaled re-trained, per factors 2, 3, and 4, with drone aero orthophoto images and is renamed to Aero-VDSR. The RGB drone orthophoto of spatial resolution 5cm/pixel (Fig. 4) has been utilized to create the aero image dataset for our experiment. It has been generated by Map Ltd employing a standard photogrammetric approach, based on vertical overlapping RGB images acquired by the drone eBee plus.

The aero orthophoto dataset contains 3,388 images of pixels size 227×227. Each image was cut into 6 patches of 41×41 pixels, thus totally 20,328 observation patches were obtained and used for VDSR re-training.

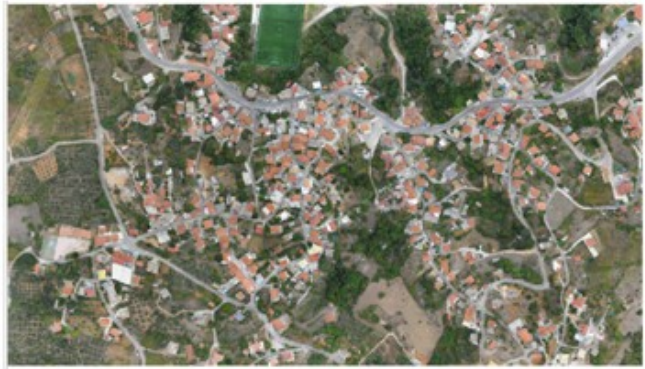

**Fig. 4.** The drone aero orthophoto of 5cm/pixel that served for creating the aero orthophoto dataset.

## 4. Experimentation

### 4.1. Settings

In the present work, deep learning trainings were carried out and in Table I the training parameter values for the different network methodologies are given. A mini batch size equal to 64 was selected as the most appropriate after experimentation with several different values. Regarding the learning rate, it was fixed at the value 0.1.

With respect to the learning rate factor, all convolutional layers had bias and weight factor equal to 1.

In the case of the -VDSR(MSE) methodology, 8 training epochs were chosen, while in the -VDSR(Var-norm) methodology the training epochs equaled 5. Training time is an important issue in deep learning. The networks with the Var-norm estimator in the regression layer exhibit a faster learnability than those which have MSE as loss function. Thereafter, the number of training epochs has been chosen so that to have a balance between good achieved generalization ability and fast training. A summary comparison of the networks is given in Table II.

TABLE I. TRAINING PARAMETER VALUES

| Metho-dology | Parameter | | |
|---|---|---|---|
| | *Epochs Number* | *Time* | *RMSE/LOSS* |
| RS-VDSR (MSE) | 8 | 15h46min | 0.53/0.1 |
| RS-VDSR (Var-norm) | 5 | 9h29min | $0.62/7.7 \times 10^{-5}$ |
| Aero-VDSR (MSE) | 8 | 21h8min | 0.60/0.20 |
| Aero-VDSR (Var-norm) | 5 | 14h4min | $0.74/1.1 \times 10^{-4}$ |

TABLE II. SUMMARY COMPARISON OF THE NETWORKS

| Characte-ristic | Network | | |
|---|---|---|---|
| | *RS-VDSR* | *Aero-VDSR* | *VDSR* |
| Estimator in regression layer during training | MSE or Var-norm | MSE or Var-norm | MSE |
| Estimator in regression layer during prediction | MSE or Var-norm | MSE or Var-norm | MSE |
| Trained with Sentinel-2 Demokritos patches | YES | NO | NO |
| Trained with drone aero orthophoto patches | NO | YES | NO |
| Trained with natural patches | YES | YES | YES |

## 4.2. Predictions

Regarding RS-VDSR and VDSR, prediction tests were performed with whole images from the Demokritos dataset. Synthesized experiments for spatial resolution enhancement per factor of 4 were carried out. The original image of size 227×227 pixels was downsampled to the size 56×56 pixels. Then, the network was asked to super-resolve the downsampled image to the original image size of 227×227 pixels. Six different scenes from Demokritos dataset were selected for the prediction tests. Fig. 5 depicts the net output or residual image, that is the estimated difference between the bicubic interpolated and the desired HR luminance components, as far as the networks VDSR, RS-VDSR(MSE) and RS-VDSR(Var-norm) are concerned. The corresponding SR luminance components of the images in the Ycbcr color space are shown in Fig. 6.

The numerical comparison results are given in Table III. Values of PSNR in dB and of the structural similarity index (SSIM) have been calculated to compare the original image luminance component with the SR luminance components that are obtained by the networks VDSR, RS-VDSR(MSE) and RS-VDSR(Var-norm). Both net methodologies RS-VDSR(MSE) and RS-VDSR(Var-norm) outperform the VDSR net. In fact, RS-VDSR(Var-norm) could be preferred over RS-VDSR(MSE) due to being faster.

During the Aero-VDSR and VDSR prediction tests, whole images from the aero orthophoto dataset were utilized. Synthesized experiments with resolution enhancement factor equal to 4 were performed. The network was asked to super-resolve the downsampled image of size 56x56 pixels to the original image size of 227x227 pixels. The prediction tests were carried out with six different scenes from the drone aero dataset. Fig. 7 demonstrates the net outputs or residual images, regarding the networks VDSR, Aero-VDSR(MSE) and Aero-VDSR(Var-norm). The corresponding SR luminance components of the images in the Ycbcr color space are depicted in Fig. 8. Concerning the numerical comparison results, these are given in Table IV. PSNR and SSIM values have been calculated to compare the original image luminance component with the SR luminance components which are obtained by the networks VDSR, Aero-VDSR(MSE) and Aero-VDSR(Var-norm). Aero-VDSR learnability seems to be slower than that of RS-VDSR. Natural images, where VDSR has been already trained on, resemble more with aerial images than with Sentinel-2 images. The above could reason the slower learnability which is observed for Aero-VDSR. Aero-VDSR(Var-norm) could be preferred over RS-VDSR(MSE) due to being faster.

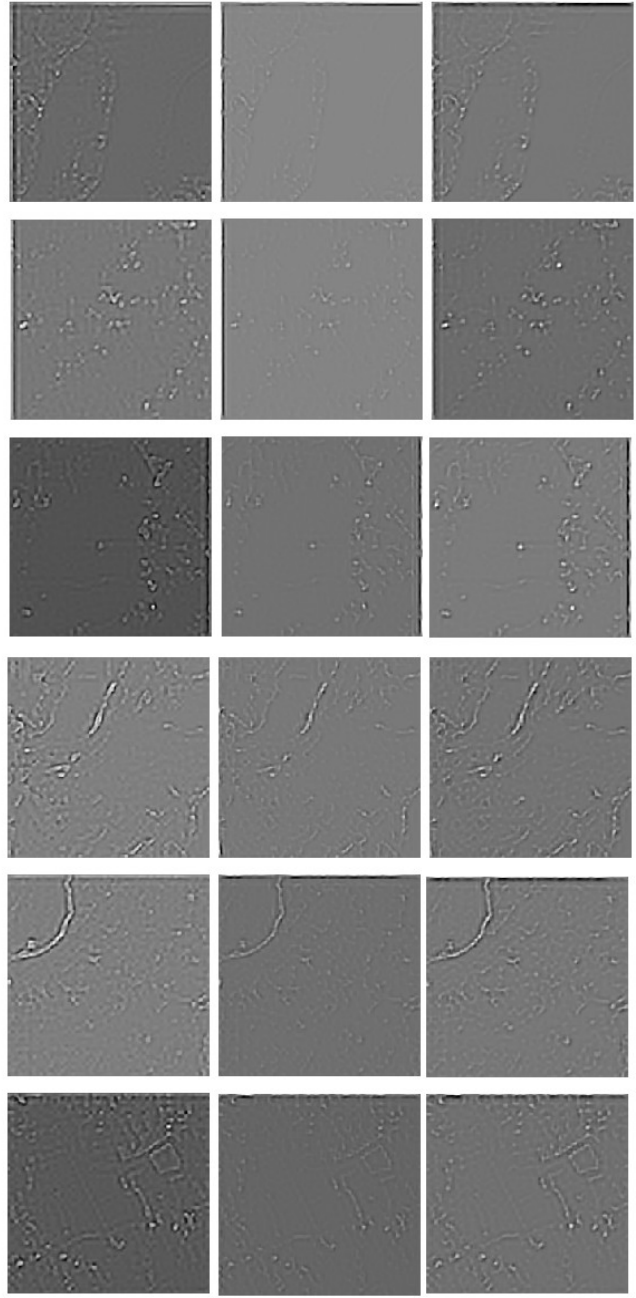

**Fig. 5.** The net output or residual image, that is the estimated difference between the bicubic interpolated and the desired HR luminance components. Rows 1 to 6 correspond to the scenes 1, 85, 292, 900, 1500, 2610 of the Demokritos dataset. The first column depicts VDSR output. The second and third columns show RS-VDSR(MSE) and RS-VDSR(Var-norm) outputs, correspondingly.

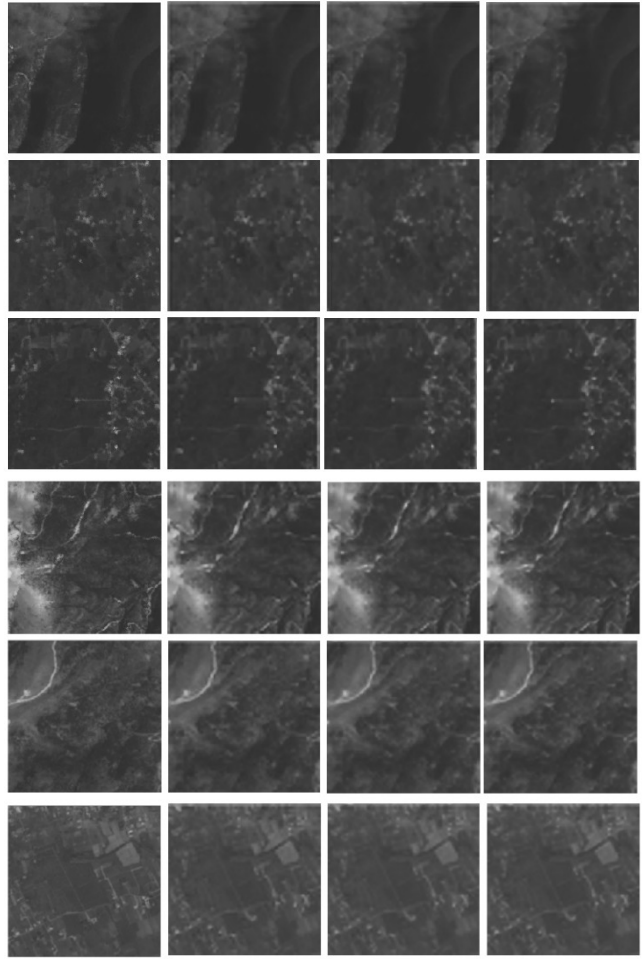

**Fig. 6.** The luminance component of the images in the Ycbcr color space. Rows 1 to 6 correspond to the scenes 1, 85, 292, 900, 1500, 2610 of the Demokritos dataset. The first column depicts the luminance component of the original image. The second, third, and fourth columns show the luminance components which are obtained by means of VDSR, RS-VDSR(MSE), and RS-VDSR(Var-norm), respectively.

TABLE III SUPER-RESOLVED VS ORIGINAL LUMINANCE COMPONENT (PSNR (DECIBEL)/SSIM) IN CASE OF SENTINEL-2 DATASET

| Demokritos Data Scene | Network | | |
|---|---|---|---|
| | *VDSR* | *RS-VDSR (MSE)* | *RS-VDSR (Var-norm)* |
| 1 | 29.55/0.916 | 32.54/0.940 | 30.90/0.930 |
| 85 | 31.16/0.917 | 34.32/0.935 | 32.39/0.930 |
| 292 | 30.55/0.909 | 32.95/0.922 | 32.23/0.923 |
| 900 | 31.05/0.811 | 31.08/0.809 | 31.31/0.817 |
| 1500 | 31.64/0.897 | 33.35/0.906 | 32.97/0.910 |
| 2610 | 32.56/0.896 | 33.43/0.902 | 33.40/0.906 |

TABLE IV SUPER-RESOLVED VS ORIGINAL LUMINANCE COMPONENT (PSNR (DECIBEL)/SSIM) IN CASE OF DRONE AERO ORTHOPHOTO DATASET

| Drone Aero Data Scene | Network | | |
|---|---|---|---|
| | *VDSR* | *Aero-VDSR (MSE)* | *Aero-VDSR (Var-norm)* |
| 1520 | 31.82/0.920 | 32.02/0.924 | 32.07/0.923 |
| 1000 | 35.14/0.922 | 35.20/0.924 | 35.24/0.924 |
| 476 | 32.16/0.866 | 32.65/0.871 | 32.59/0.869 |
| 20 | 35.52/0.908 | 35.55/0.910 | 35.52/0.910 |
| 2854 | 30.96/0.896 | 31.16/0.897 | 31.23/0.895 |
| 3354 | 32.95/0.860 | 33.06/0.863 | 32.94/0.862 |

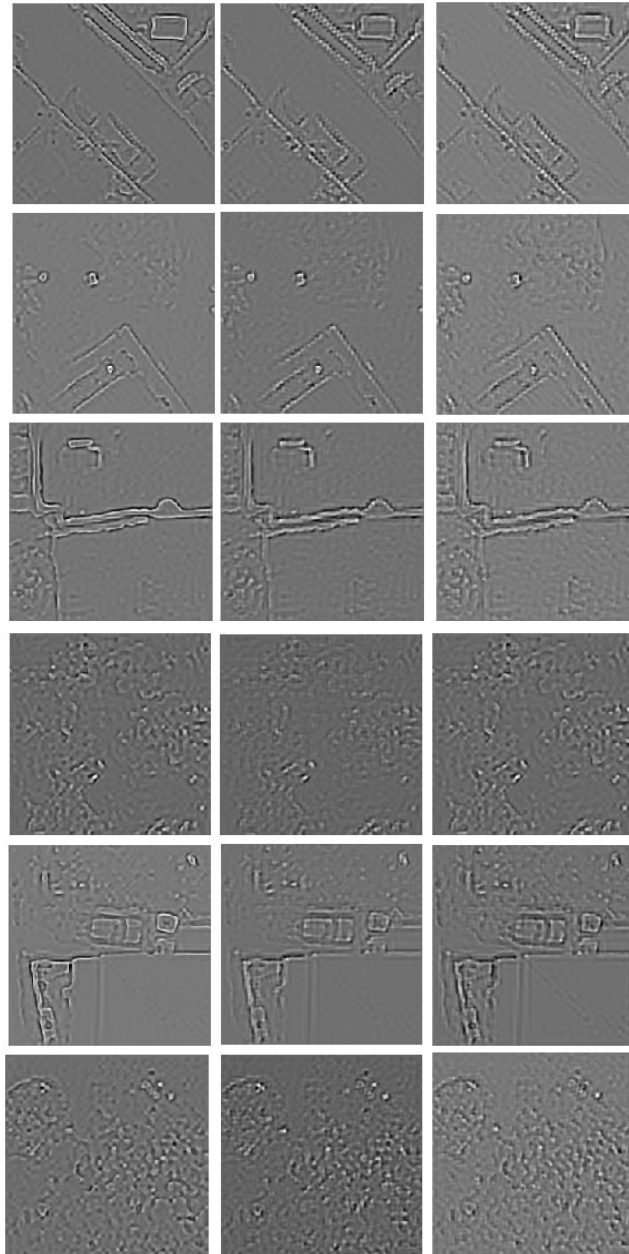

**Fig. 7.** The net output or residual image, that is the estimated difference between the bicubic interpolated and the desired HR luminance components. Rows 1 to 6 correspond to the scenes 1520, 1000, 476, 20, 2854, 3354 of drone aero dataset. The first column depicts VDSR output. The second and third columns show Aero-VDSR(MSE) and Aero-VDSR(Var-norm) outputs, correspondingly.

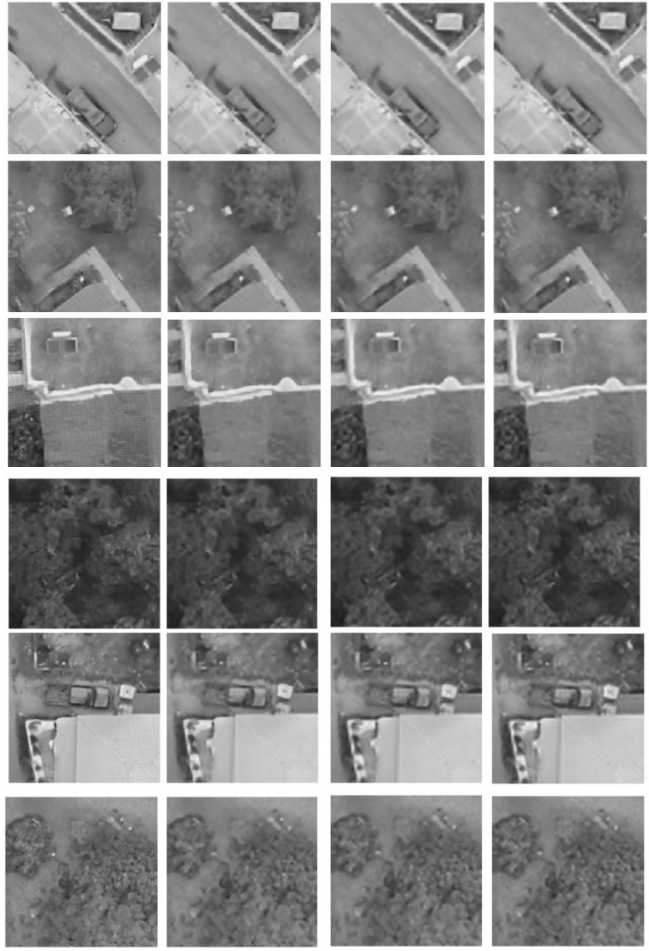

**Fig. 8.** The luminance component of the images in the Ycbcr color space. Rows 1 to 6 correspond to the scenes 1520, 1000, 476, 20, 2854, 3354 of drone aero dataset. The first column depicts the luminance component of the original image. The second, third, and fourth columns show the luminance components which are obtained by means of VDSR, Aero-VDSR(MSE), and Aero-VDSR(Var-norm), respectively.

## 5. Conclusions

In the present work, the VDSR technique is applied for improving the spatial resolution of RS images per a factor of 4. The VDSR net is separately re-trained with Sentinel-2 images and with drone aero orthophoto images. Then, the networks RS-VDSR and Aero-VDSR, respectively, are obtained. The Var-norm estimator, in the regression layer of the convolutional neural network during re-training and prediction, is proposed as a novel loss function. According to numerical and optical comparisons, the proposed nets RS-VDSR and Aero-VDSR predominate over VDSR during prediction with RS images.

For future work, predictions with images completely unknown to the net are planned. Other interesting tests could be carried out with image content loss or texture loss as network loss function, since in the present study only pixel loss was considered. Additionally, experimentation with noisy aerial images could further help show the rival behavior of MSE and Var-norm estimators during outliers rejection.